# Cable bacteria as long-range biological semiconductors

Robin Bonné [1], Ji-Ling Hou[1], Jeroen Hustings[1], Mathijs Meert[1], Silvia Hidalgo-Martinez [2], Rob Cornelissen[1], Jan D'Haen[3], Sofie Thijs[4], Jaco Vangronsveld[4,5], Roland Valcke[6], Bart Cleuren[7], Filip J. R. Meysman[2,8] and Jean V. Manca[1]*

## Abstract

*Filamentous cable bacteria exhibit unprecedented long-range biological electron transport[1–3], which takes place in a parallel fibre structure that shows an extraordinary electrical conductivity[4] for a biological material. Still, the underlying electron transport mechanism remains undisclosed. Here we determine the intrinsic electrical properties of individual cable bacterium filaments. We retrieve an equivalent electrical circuit model, characterising cable bacteria as resistive biological wires. Temperature dependent experiments reveal that the charge transport is thermally activated, and can be described with an Arrhenius-type relation over a broad temperature range (-196°C to +50°C), thus excluding metal-like electron transport. Furthermore, when cable bacterium filaments are utilized as the channel in a field-effect transistor, they show n-type transport, indicating that electrons rather than holes are the charge carriers. Electron mobilities are in the order of $10^{-1}$ cm²/Vs, comparable to many organic semiconductors[5]. This new type of biological centimetre-range semiconductor with low resistivity offers new perspectives for both fundamental studies and applications in (bio)electronics.*

*Keywords*: Cable bacteria, long-range electron transport, conductivity, activation energy, mobility, impedance spectroscopy, field-effect transistor, bioelectronics, organic electronics

[1] X-LAB, Hasselt University, Agoralaan D, B-3590 Diepenbeek, Belgium
[2] Department of Biology, University of Antwerp, Universiteitsplein 1, B-2610 Wilrijk, Belgium
[3] Institute for Materials Research (IMO-IMOMEC), Hasselt University, Wetenschapspark 1, B-3590 Diepenbeek, Belgium
[4] Centre for Environmental Sciences, Hasselt University, Agoralaan D, B-3590 Diepenbeek, Belgium
[5] Department of Plant Physiology, Faculty of Biology and Biotechnology, Maria Curie-Sklodowska University, Plac Marii Skłodowskiej-Curie 5, 20-400 Lublin, Poland
[6] Molecular and Physical Plant Physiology, Hasselt University, Agoralaan D, B-3590 Diepenbeek, Belgium
[7] Theoretical Physics, Hasselt University, Agoralaan D, B-3590 Diepenbeek, Belgium
[8] Department of Biotechnology, Delft University of Technology, Van der Maasweg 9, 2629HZ Delft, The Netherlands.
* Corresponding author: jean.manca@uhasselt.be

In 2012, a novel group of filamentous bacteria was discovered[1], which thrive in marine and freshwater sediments[6–8]. From the analysis of the sediment chemistry, it was proposed that they have the ability to transport electric currents over centimetre distances[1,9]. These so-called cable bacteria (CB) form an unbranched filamentous chain of cells that vertically align in the sediment to take advantage of the redox gradients that occur in natural sediment[6]. Metabolic oxidation and reduction half-reactions occur in different parts of the filament, and to ensure the electrical coupling of these reactions, electrons are transported over centimetre-scale distances along the filament[1].

Direct electrode measurements reveal that individual cable bacterium filaments can guide nanoampere currents over distances up to 1 centimetre under an externally applied potential[4]. This length-scale of conduction greatly surpasses that of other known current-producing bacteria[10,11], such as *Geobacter sulfurreducens* and *Shewanella oneidensis* MR-1, which act as model organisms in the fields of electromicrobiology and form conductive nanowires that are a few micrometre long. The conductive structures that enable the long-range transport in cable bacteria have recently been disclosed. Microscopy investigations reveal that all cells within a cable bacterium filament share a common space within the cell envelope, and that a network of parallel fibres run within this periplasmic space along the whole filament[12,13]. These fibres were confirmed as the primary conductive structures inside cable bacteria[4] and their conductivity exceeds 10 S/cm (Figure 1a). Yet, the underlying electron transport mechanism remains currently undisclosed. To gain insight into this electron transport, we have determined the intrinsic electrical properties of individual isolated cable bacterium filaments using a variety of electrical characterisation techniques.

Firstly, we applied electrical impedance spectroscopy to dried individual filaments and determined the representative equivalent electrical circuit. Single filaments were taken out of a sediment enrichment, and were used either as an intact filament (n=7) or as a fibre sheath (n=4) (Figure 1a). So-called "fibre sheaths" represent cable bacterium filaments from which the lipid membranes and internal cytoplasm are removed by chemical extraction, thus retaining a sheath structure that embeds the conductive fibres[4,13]. As in previous measurements[4], a single filament is placed as a connection between two electrodes (non-conductive gap size ranging from 0.1 to 1 mm) in nitrogen atmosphere to prevent current decay (Figure 1b,c). Figure 2 shows a typical impedance plot for an intact filament and a fibre sheath. All samples demonstrated similar behaviour, and gave a semicircle in the complex plane, which can be described by a R(CR) circuit: two serial resistors of which one is in parallel with a capacitor[14]. The bulk resistance of the filament is included in the parallel resistance $R_p$ whereas the contact resistance of the probe-substrate interface and the substrate-filament interface are included in the series resistance $R_s$. The capacitance $C_p$ is not intrinsically related to the cable bacterium filament, but is attributed to the stray capacitance of the measurement system, as confirmed by a reference measurement of the substrate without a filament crossing the gap[15]. $R_p$ values range from 0.7 to 3588.2 MΩ (Figure 2, Supplementary table 1), indicating that individual filaments show a variability in conductivity, as also previously observed in direct current (DC) measurements[4]. The value of $R_p$ corresponds to the total resistor value of the cable bacteria measured in a subsequent performed DC measurement, proving that the obtained conductivity values correspond with the intrinsic (bulk) electrical conductivity of the cable bacteria. Values of $R_s$ and $C_p$ highly depend on the substrate used, but the ratio $R_s/R_p = 0.002 \pm 0.004$ is consistently small, indicative of a small contact resistance. The equivalent circuit of a cable bacteria thus establishes that they can be considered as resistors, with negligible reactive impedance. This result is in line with an impedance analysis for transport in stochastic systems.[15]

The electrical conductivities of the periplasmic fibres in cable bacteria as obtained from DC and AC experiments exceed 10 S/cm, which is at the boundary region between semiconductors and metals[4,16]. To further understand the charge transport[17–19] in cable bacteria, we studied the conductivity at different temperatures for a broad temperature range in helium to avoid degradation during the measurement. Figure 3 shows the conductance $G$ as a function of the inverted thermal energy $1/k_BT$, for both an intact filament and a fibre sheath, when cooled down in discrete steps from +50 °C to -196 °C. Both filament types demonstrate a similar behaviour; the conductance decreases with decreasing temperature, thereby excluding the possibility of a metal-like conduction, as has been previously proposed for the transport mechanism in *G. sulfurreducens*[20,21]. The activation energy $E_a$ is determined by fitting the data with the Arrhenius function $G = G_0 \exp(-E_a/k_B T)$ [22] (Figure 3). The fitted curves show similar slopes, indicating comparable activation energies - the difference in offsets indicate a different room temperature conductance as observed above. Heating the samples back from -196 °C to +50 °C resulted in a similar activation energy, thereby proving any decay to be small (Supplementary Figure 1). An average of the activation energy for all samples (Supplementary Table-2) results in 45.5 ± 3.7 meV for intact filaments (n = 4) and 39.8 ± 6.1 meV for fibre sheaths (n = 2) – very close to the room temperature $k_B T$ value of 25 meV. This result demonstrates that electron transport in cable bacteria is thermally activated and can be described by an Arrhenius-type relation in a broad temperature range (-196 °C to +50 °C), extending far beyond the natural physiological temperature range.

To further reveal the potential semiconducting properties of cable bacteria, we examine their behaviour in a field-effect transistor (FET) configuration (see Figure 1b). In the bottom-gate bottom-contact FET configuration, a single filament is placed across source (S) and drain (D) electrodes separated by various channel length (100 µm to 300 µm) on top of a silicon dioxide/n-doped silicon gate (G) substrate. Transfer curves for an intact filament and a fibre sheath are shown in Figure 4a. Here, $I_D$, $V_{GS}$, and $V_{DS}$ represent the drain current, gate-to-source voltage, and drain-to-source voltage, respectively. At zero gate bias ($V_{GS}$ = 0) and $V_{DS}$ = 0.1 V, the sample shows a high off-state $I_D$. With increasing positive gate bias ($V_{GS}$ > 0) at 1 V/s (see Supplementary Figure 2), $I_D$ slightly increases (about 9 % at $V_{GS}$ = +80 V). In contrast, at $V_{GS}$ = -80 V, $I_D$ decreases with 9 %. This indicates that the charge density at the interface between cable bacterium and dielectric increases with increasing gate voltage, consistent with a n-type semiconductor behaviour. The output characteristics ($I_D$ versus $V_{DS}$) were determined for $V_{GS}$ varying from -30 V to +30 V. A typical characteristic can be found in Figure 4b, where the gate bias modulates the linear slope ($\partial I_D / \partial V_{DS}$) of the output characteristic. The conductivity linearly increases with gate bias $V_{GS}$ (Supplementary Figure 3), yielding a modulation rate of 3 mS/cm per volt.

Given the bias condition ($V_{DS} \ll V_{GS}$), the cable bacterium operates in the linear regime, as shown in **Error! Reference source not found.**a. The mobility of the electrons can be calculated by using the linear mode bias condition, $\mu = (\partial I_D / \partial V_{GS}) \cdot l / (w\, V_{DS} C_i)$, where $l$ is the channel length and $w$ the channel width[23], which corresponds to the width of the cable bacterium. $C_i$ is the gate capacitance per unit area and can be calculated as $C_i = \varepsilon_r \varepsilon_0 / d$, with $d$ the oxide thickness, $\varepsilon_0$ the vacuum permittivity and $\varepsilon_r$ the dielectric constant of the gate insulator. Four fibre sheaths were measured on a substrate with channel lengths ($L$) ranging from 100 to 300 µm. The resulting electron mobility (Supplementary Table 3) is in the range of 0.09 – 0.27 cm²/Vs, comparable to those of many organic small molecules and polymer thin films[5].

We have demonstrated with electrical impedance spectroscopy that cable bacteria can be considered as biological electrical wires. The conductive fibres in the cell envelope of cable bacteria display a large electrical conductivity (> 10 S/cm) and FET-mobility ($10^{-1}$ cm$^2$/Vs), which are similar in magnitude to synthetic organic semiconductors, and so the fibres qualify as long-range 'biological semiconductors'. Our temperature experiments reveal that the charge transport is thermally activated, described with an Arrhenius relation in a broad temperature range (-196°C to +50°C). The obtained activation energy is around 40 meV for both intact filaments and fibre sheaths. Since we have measured the reported electrical properties on isolated single cable bacteria filaments out of their wet natural aquatic habitat, we consider them as intrinsic material properties and hypothesize that they persist also in under in-vivo conditions.

The idea of cable bacteria as biological long-range semiconductors offers new perspectives for both fundamental studies and technological applications. Future studies should further elucidate the underlying electrical transport mechanism and the composition of the conductive fibres, in order to better understand their role in the overall metabolism of the cable bacteria, but also to explore their potential use in technological applications. From a materials perspective, the existence of a highly conductive organic material is of great interest towards the development of biocompatible, biodegradable and low cost electrical materials. Our proof-of-principle that the cable bacteria can function as active electrical channels in FETs – the building blocks of advanced electrical circuits – shows that they can be envisioned as base materials for the emerging field of bioelectronics, including visionary technologies such as biodegradable electronics. The reported intrinsic electrical properties together with the long-range electron transport and the wide temperature range of operation, are unique assets to envisage cable bacteria for future electronic applications.

# Methods

## Sample preparation

Cable bacteria were enriched in natural sediment cores incubated in oxygenated seawater, as described previously[24]. Sediment was collected at Rattekaai (Oosterschelde, The Netherlands). Single filaments of cable bacteria were picked from the sediment enrichment as described previously[13]. Filaments were washed (6x times) in MilliQ water to remove sediment debris thus providing so-called "intact filaments". Any excess of water was removed by a pipette and the sample was left to dry. Overall, about 5 min passed between the picking of a filament and the start of the current measurement. Alternatively, after washes with MilliQ, filament were exposed to a sequential extraction procedure, thus removing the cytoplasm and membranes, as described previously[13]. This provided so-called fibre sheaths. After about 45 minutes, the sample was transferred onto the electrode substrate.

## AC/DC Electrical measurements

For all electrical measurements, the substrate was placed at a probe station with two needle probes connecting to the two electrodes. In the Direct Current measurements, the probe station was connected to a Keithley 2450A sourcemeter (Keithley, USA) with triax cables, driven by the multi-tool control software SweepMe, as described in Meysman et al.[4]. For AC impedance measurements, the sample is probed with a VersaSTAT3F potentiostat (Ametek, USA), allowing impedance measurements in the range 1 MHz to 100 mHz. Data fitting was done with the ZSimpWin software package (Ametek). Conductivity $\sigma$ was calculated for all samples using $\sigma = Gl/A$, with $l$ the conduction length, $A$ the conductive area and $G = \Delta I/\Delta V$ the conductance calculated with a linear fit to the IV-diagrams.

## Field-effect transistor measurements

Field-effect measurements were done on a highly n-doped silicon wafer in a bottom-gate bottom-contact FET configuration. A 150 nm thick thermally grown silicon oxide layer served as a dielectric layer, and the bottom drain and source gold electrodes with thickness of 50 nm of the coplanar FET were defined by optical lithography to yield a channel length (L) of 100 μm. After the aforementioned washing (and extraction) treatment, an individual filaments was placed across the source and drain contacts.

The time response of the drain current upon applying a positive gate bias depends on the time to build up the conductive channel. Therefore, in order to measure consistent transfer curves, a proper gate sweeping speed was crucial. Supplementary Figure 2 shows the transfer curves measured at a gate sweeping speed varying from 10 V/s, 2 V/s, 1 V/s and 0.5 V/s. Using a fast scanning speed of 10 V/s, the signal of $I_D$ shows a large hysteresis. Stable $I_D$ is obtained by reducing the gate sweeping speed down to 2 V/s and 1 V/s.

## Temperature dependent measurements

Temperature measurements were performed with a commercial cryostat, model OptistatDN by Oxford Instruments. The cryostat is liquid nitrogen based, allowing the sample to be cooled down from ambient temperature to -196°C. Cable bacterium filaments were dropcasted on glass substrates and carbon paste was applied to both ends to form electrical contact points. These were mounted in a double-walled cryostat using spring contacts. The inner vessel was filled with helium as exchange gas; the outer vessel by a high vacuum, maintaining a pressure around $10^{-9}$ bar for thermal insulation. The sample was heated to 50°C and then cooled down to -196°C in steps of 25°C. For each step, current-voltage curves were measured to obtain the conductivity $\sigma$.

Cooling down was achieved by adjustment of a needle valve for the liquid nitrogen flow towards the heat-exchanger. A steady-state temperature is established by the combination of the heat-exchanger and an adjacent heating element. The latter was coupled in a feedback loop to a temperature control unit, model ITC 502 by Oxford Instruments. After a stabilization of the current (about 15 min), a current-voltage measurement was performed at each temperature.

# References

1. Pfeffer, C. *et al.* Filamentous bacteria transport electrons over centimetre distances. *Nature* **491**, 218–221 (2012).

2. Meysman, F. J. R. R. Cable Bacteria Take a New Breath Using Long-Distance Electricity. *Trends Microbiol.* **26**, 411–422 (2017).

3. Kjeldsen, K. U. *et al.* On the evolution and physiology of cable bacteria. *Proc. Natl. Acad. Sci.* **116**, 201903514 (2019).

4. Meysman, F. J. R. *et al.* A highly conductive fibre network enables centimetre-scale electron transport in multicellular cable bacteria. *Nat. Commun.* **10**, (2019).

5. Paterson, A. F. *et al.* Recent Progress in High-Mobility Organic Transistors: A Reality Check. *Adv. Mater.* **30**, 1–33 (2018).

6. Malkin, S. Y. *et al.* Natural occurrence of microbial sulphur oxidation by long-range electron transport in the seafloor. *ISME J.* **8**, 1843–1854 (2014).

7. Risgaard-Petersen, N., Damgaard, L. R., Revil, A. & Nielsen, L. P. Mapping electron sources and sinks in a marine biogeobattery. *J. Geophys. Res. G Biogeosciences* **119**, 1475–1486 (2014).

8. Seitaj, D. *et al.* Cable bacteria generate a firewall against euxinia in seasonally hypoxic basins. *Proc. Natl. Acad. Sci. U. S. A.* **112**, 13278–83 (2015).

9. Bjerg, J. T. *et al.* Long-distance electron transport in individual, living cable bacteria. *Proc. Natl. Acad. Sci.* **115**, 5786–5791 (2018).

10. Creasey, R. C. G. *et al.* Microbial nanowires – Electron transport and the role of synthetic analogues. *Acta Biomater.* **69**, 1–30 (2018).

11. Reguera, G. Microbial nanowires and electroactive biofilms. *FEMS Microbiol. Ecol.* **94**, 1–13 (2018).

12. Jiang, Z. *et al.* In vitro single-cell dissection revealing the interior structure of cable bacteria. *Proc. Natl. Acad. Sci.* **115**, 201807562 (2018).

13. Cornelissen, R. *et al.* The Cell Envelope Structure of Cable Bacteria. *Front. Microbiol.* **9**, 3044 (2018).

14. E. Barsoukov, J. R. M. *Impedance Spectroscopy: Theory, Experiment, and Applications*. (Wiley Subscription Services, Inc., A Wiley Company, 2005). doi:10.1002/0471716243

15. Cleuren, B. & Proesmans, K. Stochastic impedance. *Phys. A Stat. Mech. its Appl.* (2019). doi:10.1016/j.physa.2019.122789

16. Kittel, C. & Holcomb, D. F. Introduction to Solid State Physics. *Am. J. Phys.* **35**, 547–548 (1967).

17. Mott, N. F. Conduction in non-crystalline materials. *Philos. Mag.* **19**, 835–852 (1969).

18. Ing, N. L., El-Naggar, M. Y. & Hochbaum, A. I. Going the Distance: Long-Range Conductivity in Protein and Peptide Bioelectronic Materials. (2018). doi:10.1021/acs.jpcb.8b07431

19. Amdursky, N., Głowacki, E. D. & Meredith, P. Macroscale Biomolecular Electronics and Ionics. *Adv. Mater.* 1802221 (2018). doi:10.1002/adma.201802221

20. Malvankar, N. S. *et al.* Tunable metallic-like conductivity in microbial nanowire networks. *Nat. Nanotechnol.* **6**, 573–579 (2011).

21. Ing, N. L., Nusca, T. D. & Hochbaum, A. I. Geobacter sulfurreducens pili support ohmic electronic conduction in aqueous solution. *Phys. Chem. Chem. Phys.* **19**, 21791–21799 (2017).

22. Marcus, R. Electron transfer reactions in chemistry. Theory and experiment. *J. Electroanal. Chem.* **438**, 251–259 (1993).

23. Lu, W., Xie, P. & Lieber, C. M. Nanowire Transistor Performance Limits and Applications. *IEEE Trans. Electron Devices* **55**, 2859–2876 (2008).

24. Burdorf, L. D. W. *et al.* Long-distance electron transport occurs globally in marine sediments.

*Biogeosciences* **14**, 683–701 (2017).

# End notes

**Acknowledgements** The authors thank the colleagues from X-LAB from Hasselt University and the Microbial Electricity team from University of Antwerp for discussions and feedback. Special thanks to K. Ceyssens and T. Custers for the graphics and M. De Roeve for help with the experimental setup. This research was financially supported by the Research Foundation - Flanders (FWO project grant G031416N to FJRM and JM and FWO aspirant grant 1180517N to RB). FJRM was additionally supported by the Netherlands Organization for Scientific Research (VICI grant 016.VICI.170.072).

**Author Contributions** Impedance measurements were performed by R. B.; transistor measurements by R. B., JL. H. and M. M.; cryostat measurements by JL. H. and J. H. SEM imaging was done by J.D.; cable bacteria cultivation and sample preparation was done by S. HM., R. B., JL. H. and F. M. The study was conceived by J. M., F.M. and JL. H. Furthermore, R. C., S. T., J. V., R. V., F. M. and B. C. contributed to discussions and preparation of the manuscript.

**Author Information** The authors declare no competing interests. Correspondence and requests for materials should be addressed to R. B. (robin.bonne@uhasselt.be) or J. M. (jean.manca@uhasselt.be).

# Figures

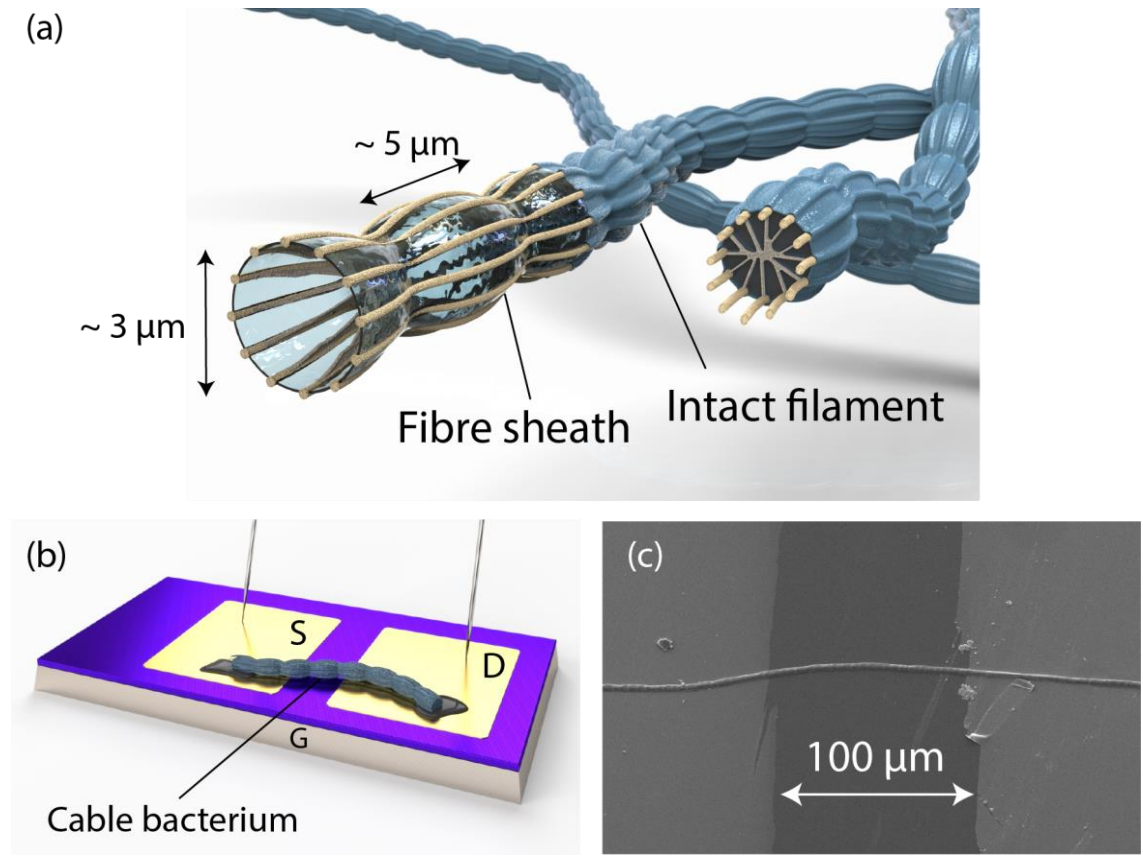

**Figure 1 | Cable bacteria and set-up for electrical measurements.** (a) A graphical representation shows a cable bacterium to be a filament of up to 10.000 cells. Samples for electrical measurements are either intact filaments, or a fibre sheath after a removal of the cytoplasm and membranes of the bacterium. (b) The filament connects two gold or carbon paste electrodes for DC or AC measurements. For FET measurements, the two gold electrodes act as source (S) and drain (D) compared to the highly n-doped silicon wafer gate (G) electrode at the bottom. (c) SEM image of a cable bacterium crossing two gold electrodes with an interdistance of 100 μm.

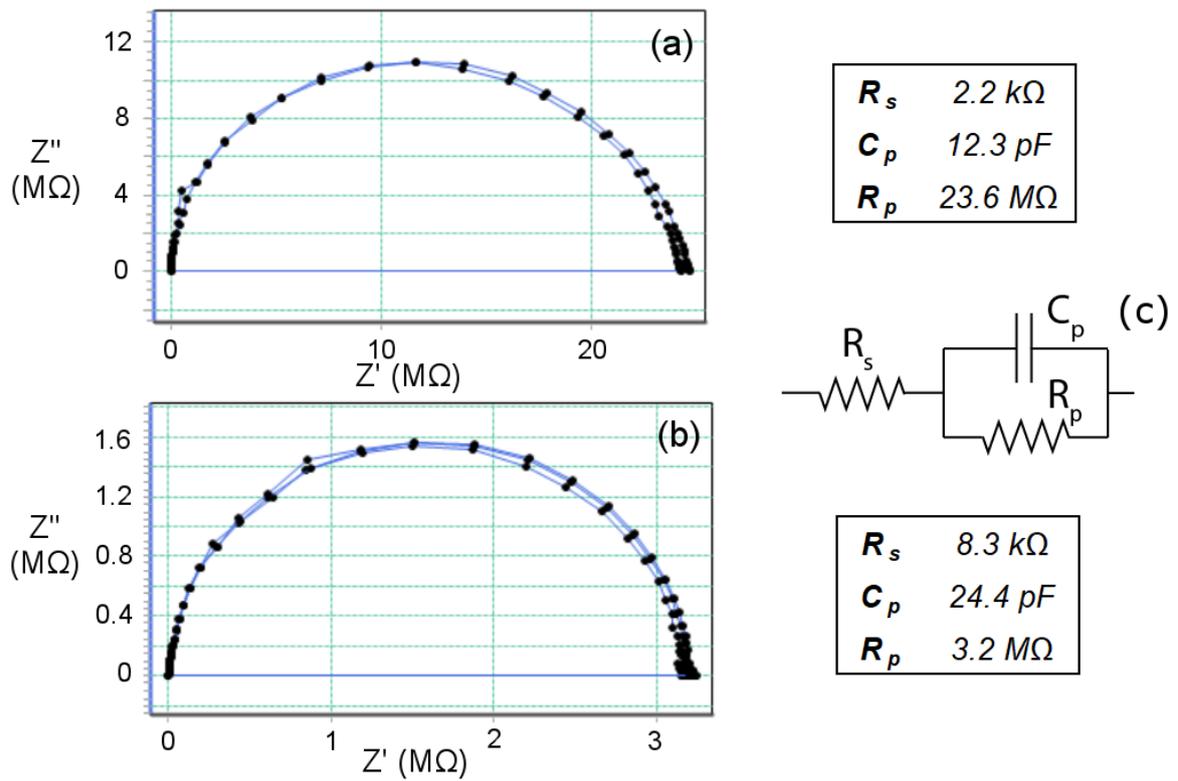

**Figure 2 | Impedance spectroscopy and equivalent circuit show cable bacteria to behave as resistors.** Impedance spectroscopy results for (a) an intact filament and (b) a fibre sheath, showing three overlapping consecutive measurements. (c) shows an R(CR)-scheme as the equivalent circuit, with resistor $R_s$ in series with a parallel stack of a capacitor $C_p$ and a resistor $R_p$.

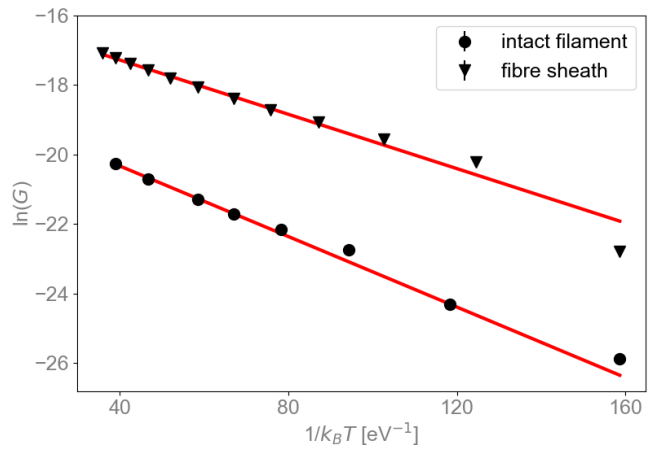

**Figure 3 | Temperature dependent experiments show conductivity described by Arrhenius-relation.** Arrhenius plots for the temperature dependent measurements of intact filaments and fibre sheaths show the conductance $G$ as a function of the inverted thermal energy $1/k_BT$.

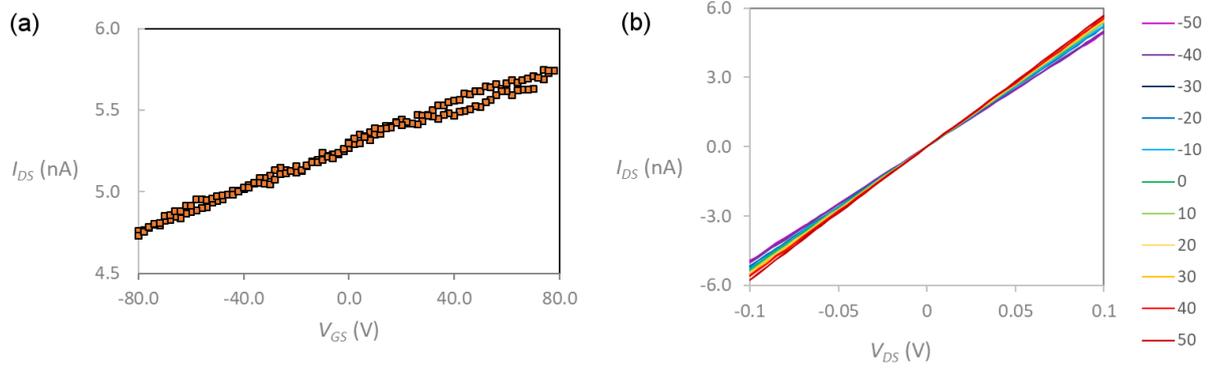

**Figure 4 |** **FET measurements reveal an n-type semiconductor behaviour for cable bacteria as channel.** (a) Transfer characteristics of a fibre sheath CB-FET measured at a constant $V_{DS}$ = 0.05 V show a slight modulation of the drain current $I_D$ when the gate bias $V_{GS}$ is changed from 0 V to 80 V to -80 V and back to 0 V (as indicated by the arrows). (b) Output characteristics of a fibre sheath CB-FET under a constant gate voltage varying from -30 V to +30 V show the slope of the current-voltage curve to change as a function of gate bias $V_{GS}$.

# Supplementary Information

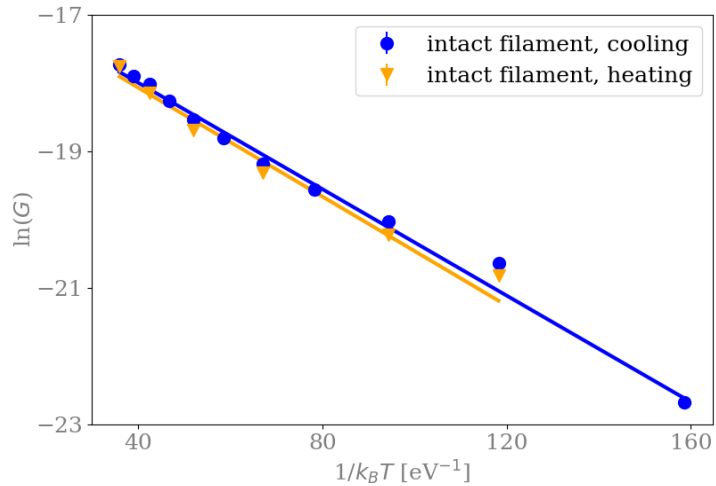

**Figure 1 |** Cooling and consequent heating of a cable bacterium filament shows the same Arrhenius behaviour with a similar activation energy and hence no decay of the sample.

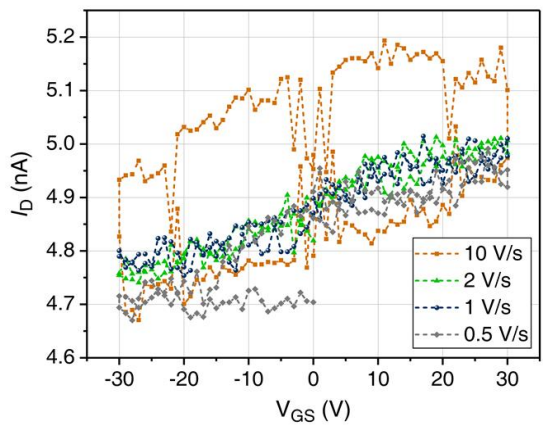

**Figure 2 |** Linear transfer characteristics of fibre sheath CB-FET measured at a gate sweeping speed varying from 10 V/s, 2 V/s, 1 V/s and 0.5 V/s. $V_{DS}$ is constant at 0.05 V.

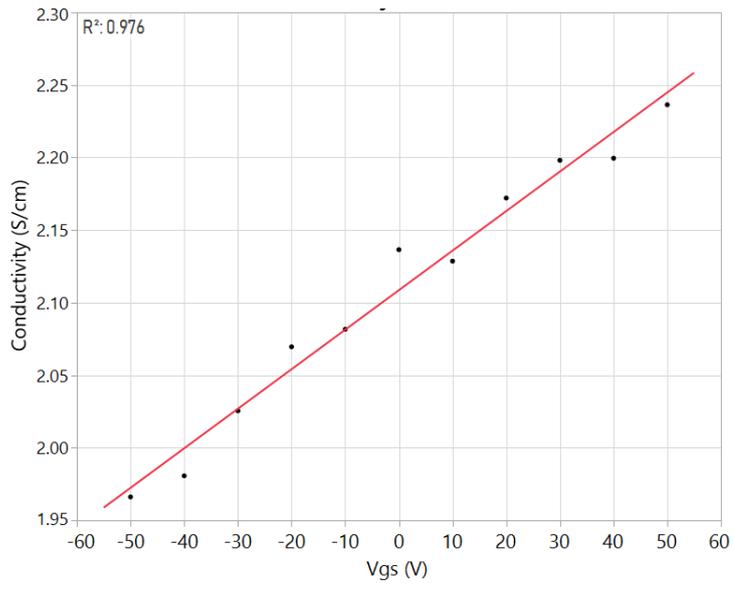

**Figure 3 |** Conductivity as a function of gate voltage $V_{GS}$ shows a modulation rate of about 3 mS cm$^{-1}$.

Table 1 | **Impedance spectroscopy (IS) fitting results show a small contact resistance.** For 7 intact filaments and 4 fibre sheaths, a fit is made with a R(RC)-circuit. Here, Rs the series resistance, Rp the parallel resistance and Cp the parallel capacitance, with $R_p$ showing a large variance in bulk conductivity values. The low value for $R_S/R_p$ indicates the contact resistance is negligible as compared to the bulk resistance.

| Sample | $R_s$ (kΩ) | $R_p$ (MΩ) | Cp (pF) | $R_s/R_p$ (%) |
|---|---|---|---|---|
| IS-Intact1 | 0.584 | 64.404 | 39.166 | 0.001 |
| IS-Intact2 | 0.601 | 9.896 | 40.398 | 0.006 |
| IS-Intact3 | 0.445 | 15.916 | 39.810 | 0.003 |
| IS-Intact4 | 0 | 1112.800 | 45.228 | 0 |
| IS-Intact5 | 102.211 | 8.138 | 1.956 | 1.256 |
| IS-Intact6 | 100.599 | 11.359 | 1.764 | 0.886 |
| IS-Intact7 | 80.635 | 150.982 | 2.103 | 0.053 |
| IS-Sheath1 | 2.467 | 26.552 | 15.056 | 0.009 |
| IS-Sheath2 | 0.575 | 0.705 | 56.708 | 0.082 |
| IS-Sheath3 | 8.458 | 3.078 | 25.912 | 0.275 |
| IS-Sheath4 | 0 | 3588.200 | 11.141 | 0 |

**Table 2 | Arrhenius analysis for temperature dependent measurements on intact filaments and fibre sheaths.** The chi-squared value χ2, conductance $G_0$ at temperature $(k_BT)^{-1} = 0$ and the activation energy $E_{act}$ are calculated based on an Arrhenius fitting on 4 intact filaments and 2 fibre sheaths.

| Sample | χ2 | $G_0$ (nS) | $E_{act}$ (meV) |
|---|---|---|---|
| T-Intact1 | 0.0141 | 11.291 ± 0.030 | 50.738 ± 0.059 |
| T-Intact2 | 0.0115 | 72.662 ± 0.050 | 38.956 ± 0.012 |
| T-Intact3 | 0.0148 | 97.040 ± 0.049 | 37.044 ± 0.009 |
| T-Intact4 | 0.1027 | 22.865 ± 0.064 | 51.740 ± 0.052 |
| T-Sheath1 | 0.0324 | 4.484 ± 0.024 | 47.222 ± 0.112 |
| T-Sheath2 | 0.0414 | 149.681 ± 0.106 | 39.078 ± 0.015 |

**Table 3 | Mobility calculated for the field effect transistor measurements on four cable bacterium sheaths.**

| Sample | Mobility μ (cm² V$^{-1}$s$^{-1}$) |
|---|---|
| FET-Sheath1 | 0.09 |
| FET-Sheath2 | 0.11 |
| FET-Sheath3 | 0.27 |
| FET-Sheath4 | 0.27 |